\documentclass[12pt]{revtex4}

\newcommand{\be}{\begin{equation}}
\newcommand{\ee}{\end{equation}}
\newcommand{\ba}{\begin{eqnarray}}
\newcommand{\ea}{\end{eqnarray}}

\usepackage{color}
\usepackage{latexsym}
\usepackage{amsfonts,amsbsy}

\def\qed{\hbox{${\vcenter{\vbox{
   \hrule height 0.4pt\hbox{\vrule width 0.4pt height 6pt
   \kern5pt\vrule width 0.4pt}\hrule height 0.4pt}}}$}}

\begin{document}
\title{Wu-Yang ambiguity in connection space}

\author{R. Aldrovandi and A. L. Barbosa} 
\affiliation{Instituto de F\'{\i}sica Te\'orica \\
 Universidade Estadual Paulista\\
Rua Pamplona 145 \\
01405-900 S\~ao Paulo SP \\ Brazil} 

\begin{abstract}

Two distinct gauge potentials can
have the same field strength, in which case they are said to be
``copies'' of each other. The consequences of this possibility 
for the general  space ${\mathcal A}$ of gauge potentials
are examined.  Any two potentials  are connected by a straight line in
${\mathcal A}$, but  a straight line going through two copies either
contains no other copy or is entirely formed by  copies.

\end{abstract}





\maketitle 

\section{Introduction}

A good understanding of the Wu-Yang ambiguity~\cite{WY75}, with all 
its aspects and consequences, does not seem to have been as yet achieved.
Activity on the subject has been intensive in the first years after its
discovery~\cite{Col77,Hal78,Sol79,BW79,DD79,Mos79} and
declined afterwards~\cite{MS01}. Progress has been  made step by step,
sometimes through the discovery of general  properties of formal
character~\cite{Dor81}, most of times by unearthing particular  cases
which elucidate special points~\cite{BGT79}.  This note is devoted to
the  presentation of one more formal property, which shows up in the  space
${\mathcal A}$ of gauge potentials, or connections (``$A$-space''). 
We shall mostly use invariant notation, such as 
$A = J_a A^a{}_\mu dx^\mu $ for gauge potentials (connections), $F = 
\frac{1}{2} J_a F^a{}_{\mu \nu} dx^\mu \wedge dx^\nu$ for the field 
strength (curvatures), $K = J_a K^a{}_\mu dx^\mu $ for covectors in 
the adjoint representation of the gauge algebra, etc.  Wedge products 
will be implicit.

Consider two connection forms $A$ and $A^{\sharp}$, with curvatures
\[
F = d A + A A \ \ \  {\textnormal{and}} \ \ \ F^{\sharp} = dA^{\sharp} +
A^{\sharp} A^{\sharp}. 
\]
Under a gauge transform given by a member of the gauge group 
represented by g in the adjoint representation, the connections will
transform  according to
\[
A \Rightarrow A' = gAg^{-1} + gdg^{-1}\ \ \ ;\ \ \ A^{\sharp} \Rightarrow 
A^{\sharp \; \prime} = gA^{\sharp}g ^{-1} + gdg^{-1}.
\]
The curvatures are 2-forms  covariant under these transformations: $F
\Rightarrow F' = gFg^{-1}$, $F^{\sharp}
\Rightarrow F^{\sharp \prime} = gF^{\sharp}g^{-1}$. The  difference 
\[K = A^{\sharp} - A 
\]
will be a 1-form transforming according to
\[
K^{\prime} = A^{\sharp 
\; \prime} - A^{\prime} = g(A^{\sharp}- A)g^{-1} = g^{-1} K g .
\]
That is, the difference between two connections is a covariant 1-form 
in the adjoint  representation. The ambiguity appears when 
$F^{\sharp} = F$ but $K \ne 0$. Gauge potentials like
$A$ and $A^{\sharp}$, corresponding to the same field strength, are
frequently called ``copies''.
 
It is well known that gauge
covariance divides ${\mathcal A}$ into equivalence classes, each class
representing a potential up to gauge transformations. The space of
gauge inequivalent connections is ${\mathcal \alpha}$ = 
${\mathcal A}/{\mathcal G}$, where ${\mathcal G}$ is the so-called large
group of point-dependent group elements \cite{{FS84},{WZ85}}. 
We shall not enter into the details of the ${\mathcal A}$-space structure 
\cite{AP95}. Let us only say that, technically, only
variations not along the large group are of interest for
copies.  This excludes gauge transformations. 

The whole point is that $F$ does not determine $A$. At each point of
spacetime  a gauge can be chosen in which $A = 0$ and consequently $F =
dA$. This is true also along a line. One might think of integrating by
the homotopy  formula~\cite{NS91} to obtain $A$ from $F$. This is
impossible because the involved homotopy requires  the validity of $F =
dA$ on a domain of the same dimension of spacetime
 and the alluded gauge cannot exist (unless $F= 0$) on a domain of
dimension $2$ or higher~\cite{ABP03a}. 
For copies, the difference form $K$
defines a translation on space  ${\mathcal A}$ leaving $F$ invariant. This
invariance establishes another division of  ${\mathcal A}$ in equivalence
classes. In effect, define the relation
$R$ by: $A R A^{\sharp}$ if $A^{\sharp}$ is a copy  of $A$.  This
relation is reflexive, transitive and symmetric, consequently an 
equivalence.  The space of connections with distinct curvatures will  be
the quotient ${\mathcal \alpha}/R$.

The gauge group element g can be seen as a matrix acting on 
column-vectors V belonging to an associated vector representation.  
The covariant differentials according to $A$ and $A^{\sharp}$ will 
have, in the vector representation, the forms $DV = dV + AV$; 
$D^{\sharp}V = dV + A^{\sharp}V = dV + (A^{\sharp} - A)V + AV$, that 
is,
\[
D^{\sharp}V = DV + K V .
\]
For a matrix 1-form like the difference 1-form K,
\[DK = dK + AK + K A = dK + \{A, K\};\]
\[
D^{\sharp} K = dK + A^{\sharp} K + K A^{\sharp} = dK + \{ A^{\sharp}, 
K\}.
\]
It is immediately found that 
\begin{equation}
D^{\sharp} K = DK + 2 K K \label{DsharpofK}
\end{equation}
and the 
relation between the two curvatures is 

\be F^{\sharp} = F + DK + KK. \label{FsharpF}
\ee 
A direct calculation gives
\be%
					DDK + [K, F] = 0,         \label{eq:Bianchidiff} 
\ee%
which actually holds for any covariant 1-form 
in the adjoint  representation.

Equation (\ref{FsharpF}) leads to a general result: given a connection $A$ defining a covariant
derivative $D_A$, each solution $K$ of  $D_AK + KK = 0$ will give a copy. 

\section{On the connection space ${\mathcal A}$}
We have been making implicit use of one main property of the space
${\mathcal A}$ of connections, namely:   ${\mathcal A}$ is a convex affine
space, homotopically trivial
\cite{Sin78}. One way to  state this operationally~\cite{Sin81}
has been used above: given a connection $A$, every
other connection $A^{\sharp}$ can be written as
$A^{\sharp} = A + K$, for some covariant covector K. Another way is:
through any two connections $A$ and $A^{\sharp}$ there exists a
straight line of connections
$A_{t}$, given by
\be %
A_{t} = t A^{\sharp} + (1-t) A.  \label{At1}
\ee %
In this expression $t$ is a real parameter, $A_{0} = A$ and $A_{1} =
A^{\sharp}$. In  terms of the difference form $K$, that straight line
is written
\be %
A_{t}   = A + t K = A^{\sharp} - (1 - t) K. \label{At2} 
\ee %
Of course, $\frac{dA_{t}}{ dt} = K$.  Indicating by $D_{t}$ the
covariant derivative according to connection $A_{t}$, we find
\begin{equation}
D_{t} K = DK + 2 t KK. \label{DtK}
\end{equation}
The curvature of $A_{t}$ is 
\be %
F_{t} = dA_{t} + A_{t} A_{t} = t F^{\sharp} + (1-t) F + t(t-1) K K ,
\ee %
or 
\be %
F_{t} = F + t DK + t^{2} K K = F + t D_t K - t^{2} K K. \label{Ft}
\ee %
Notice $F_{0} = F$, $F_{1} = F^{\sharp}$.  It
follows that 
\begin{equation}
\frac{dF_{t}}{dt} = DK + 2 t KK = D_{t} K . \label{dFdt}
\end{equation}
\section{The copy-structure of space ${\mathcal A}$}

The results of the previous section are valid for any two connections
$A$, $A^{\sharp}$.  Let us address the question of copies.  
From (\ref{FsharpF}), the necessary and sufficient condition to have
$F^{\sharp} = F$ is
\be %
DK + KK = 0.  \label{eq:necond}
\ee %
From
the Bianchi identity $D^{\sharp}F^{\sharp} = 0$ applied with
$A^{\sharp} = A + K$  it follows that
\be %
[K, F] = 0.  \label{KFcomm}
\ee %
These conditions~\cite{DW76} lead to the well-known determinantal
conditions 
\cite{Ros77}, \cite{Cal77} for 
the non-existence of copies. Notice that (\ref{eq:Bianchidiff}) and
(\ref{KFcomm})  imply $DD K = 0$. Copies are of interest only for
non-abelian theories. In the abelian case $KK \equiv 0$, $DK \equiv dK$, and
condition (\ref{eq:necond}) reduces to $dK = 0$, which means that locally $K
= d \phi$ for some $\phi$. Then $A^{\sharp} = A + d \phi$, a mere gauge
transformation.

A first consequence of the conditions above is
\[
\frac{dF_{t}}{dt}  =  D_{t} K = (1 - 2t) D K = (2t-1) KK.
\]
A second consequence is that now the line through $F$ and
$F^{\sharp}$  takes the form  
\be 
		F_{t} = F + t(t-1) K K = F + t(1-t) DK. \label{FtF}
\ee

We have thus the curvatures of all the connections linking two copies 
along a line in connection space.  Are there other copies on this line 
?  In other words, is there any $s \neq 0,1$ for which $F_{s} = F$ ?  The
existence of  one such copy would imply, by the two expressions in
Eq.(\ref{FtF}),
$DK = 0$ and $K K = 0$.  But then, by the first equality of Eq.(\ref{Ft}),
all points on the  line
$A_{t} = A + t K$ are copies.  Three colinear copies imply that
$A_{t}$ is a line entirely formed of copies.

As $DK = 0$ implies $K K = 0$ by (\ref{eq:necond}), it also implies
$D^{\sharp}K = 0$ [by (\ref{DsharpofK})] and vice-versa. Consequently, 
\begin{quotation}
 every point of the  line $A_{t} = t A^{\sharp} + (1-t) A$ through two
copies $A$ and $A^{\sharp}$ represents a copy when the  difference tensor
$K = A^{\sharp} - A$  is parallel-transported by 
either  $A$ or $A^{\sharp}$.  
\end{quotation}

In this case $\frac{dF_{t}}{dt} = 0$, that is, 
$F_{t} = F$ for all values of $t$. 
Also, $D_{t}K = 0$ for all $t$, so that $K$ is parallel-transported by
each connection on the line. Notice that an arbitrary finite $K$ such that
$DK = 0$ does not necessarily engender a line of  copies.  It is necessary
that $K$ be {\it a priori} the difference  between two copies. 
 
The  above condition is necessary and sufficient: if $F_{t} \neq F$ for 
some $t \neq 0, 1$, Eqs.(\ref{FtF}) imply both  $DK \neq 0$ and  $K K \neq
0$.  If the line joining  two copies includes one point which is not a copy,
then all other  points for $t \neq 0, 1$ correspond to non-copies.

\begin{quotation}
{\it {Given two copies and the straight line joining them, either there is 
no other copy on the line or every point of the line represents a 
copy.}}	
\end{quotation}

As a consequence, if there are copies for a certain $F$, and one of 
them (say, $A$) is isolated, then there are no copies on the lines 
joining $A$ to the other copies.  Notice, however, that the existence 
of families of copies dependent on continuous  parameters is known 
\cite{FK94}. Thus, certainly not every copy is isolated. 

The question of isolated copies is better understood by considering, instead
of the above finite $K$, infinitesimal translations on ${\mathcal A}$. In
effect,  consider the variation of $F$, $\delta F = d \delta A +
\delta A A + A \delta A = D_A \delta A$. In order to
have $\delta F = 0$ it is enough that $D_A \delta A = 0$. Consequently, 
no copy is completely isolated. There can be copies close to any $A$:   
each variation satisfying $D_A \delta A = 0$ leads to
a copy. Taken together with what has been said above on the finite case,
this means that there will be lines of copies along the ``directions'' of the 
parallel-transported
$\delta A$'s. 

 Notice that a line through copies of the vacuum is
necessarily a line of copies. In effect, given $A$ and $A^{\sharp}$ with
$F = F^{\sharp} = 0$, there is a gauge in which $A = 0$ and another gauge
in which $A^{\sharp} = 0$. Using the first of these gauges, $A_t = t
K$ along the line.  On the other hand,  $F_t = t (t-1) KK = 0$ by
(\ref{FtF}).  As
$DK + KK = 0$, we can write
$KK = DK + KK + KK = dK + AK + KA + KK + KK = dK + A^{\sharp}K +
KA^{\sharp} = D^{\sharp} K = D^{\sharp}A^{\sharp} = F^{\sharp} = 0$.
It follows that $F_t = 0$.

Summing up, the overall picture is the following: from any $A$ will emerge
lines of  three kinds:
\begin{itemize}

\item lines of copies, given by those $\delta A$ which are 
 parallel-transported  by $A$; 
\item lines of non-copies, given by those $\delta A$ which are 
not parallel-transported by $A$;

\item lines along covariant matrix 1-forms $K$ satisfying $D_AK + KK = 0$,
which will meet one copy at $A + K$, and only that one.
\end{itemize}

\section*{Acknowledgments} 
The authors are thankful to FAPESP-Brazil and CNPq-Brazil for financial
support.


\end{document}